\documentclass[12pt]{spieman}  

\usepackage{algpseudocode}
\usepackage{algorithm}
\usepackage{amsmath}
\usepackage{amsfonts}
\usepackage{multirow}
\usepackage{ntheorem}
\usepackage{threeparttable}
\usepackage{color}
\usepackage{cite}
\usepackage{graphicx}
\usepackage{setspace}
\usepackage{tocloft}
\usepackage{url}
\graphicspath{{./figs/}}

\theoremstyle{plain}
\theoremsymbol{\ensuremath{\clubsuit}}
\theoremseparator{.}
\newtheorem{myproblem}{Problem}

\cftpagenumbersoff{figure}
\cftpagenumbersoff{table} 
\begin{document}

\title{
  Triple Patterning Lithography (TPL) Layout Decomposition using End-Cutting
}

\author{
  Bei Yu\supscr{$a$}
  \footnote{The preliminary version of this work appeared in Proc. SPIE 8684, Design for Manufacturability through Design-Process Integration VII.},
  \ \ Subhendu Roy\supscr{$a$}, \ \ Jhih-Rong Gao\supscr{$b$}, \ \ David Z. Pan\supscr{$a$}
}
\affiliation{
\supscrsm{$a$} ECE Department, University of Texas at Austin, Austin, Texas, United States 78712\\
\supscrsm{$b$} Cadence Design Systems, 12515-7 Research Boulevard, Austin, Texas, United States 78759
}

\maketitle 

\begin{abstract}
Triple patterning lithography (TPL) is one of the most promising techniques in the 14nm logic node and beyond.
Conventional LELELE type TPL technology suffers from native conflict and overlapping problems.
Recently, as an alternative process, triple patterning lithography with end cutting (LELE-EC) was proposed to overcome the limitations of LELELE manufacturing.
In LELE-EC process the first two masks are LELE type double patterning,
while the third mask is used to generate the end-cuts.
Although the layout decomposition problem for LELELE has been well-studied in the literature,
only few attempts have been made to address the LELE-EC layout decomposition problem.
In this paper we propose the comprehensive study for LELE-EC layout decomposition.
Conflict graph and end-cut graph are constructed to extract all the geometrical relationships of both input layout and end-cut candidates.
Based on these graphs, integer linear programming (ILP) is formulated to minimize the conflict number and the stitch number.
The experimental results demonstrate the effectiveness of the proposed algorithms.
\end{abstract}


{\noindent
\footnotesize
{\bf Address all correspondence to}:
Bei Yu,
E-mail:  \linkable{bei@cerc.utexas.edu}
}

\begin{spacing}{1}   

\section{Introduction}
\label{sec:tplec_intro}

As the semiconductor process further scales down to the 14 nanometer logic node and beyond, the industry encounters many lithography-related issues.
Triple patterning lithography (TPL) \cite{TPL_SPIE2012_Lucas} is one of the most promising candidates as next generation lithography techniques,
along with extreme ultraviolet lithography (EUVL), electron beam lithography (EBL), directed self-assembly (DSA), and nanoimprint lithography (NIL)
\cite{LITH_TCAD2013_Pan,LITH_ICCAD2012_Yu}.
However, all these new techniques are challenged by some problems, such as tremendous technical barriers, or low throughput.
Therefore, TPL earns more attentions recently from both industry and academia.

\begin{figure}[tb]
  \centering
  \includegraphics[width=0.70\textwidth]{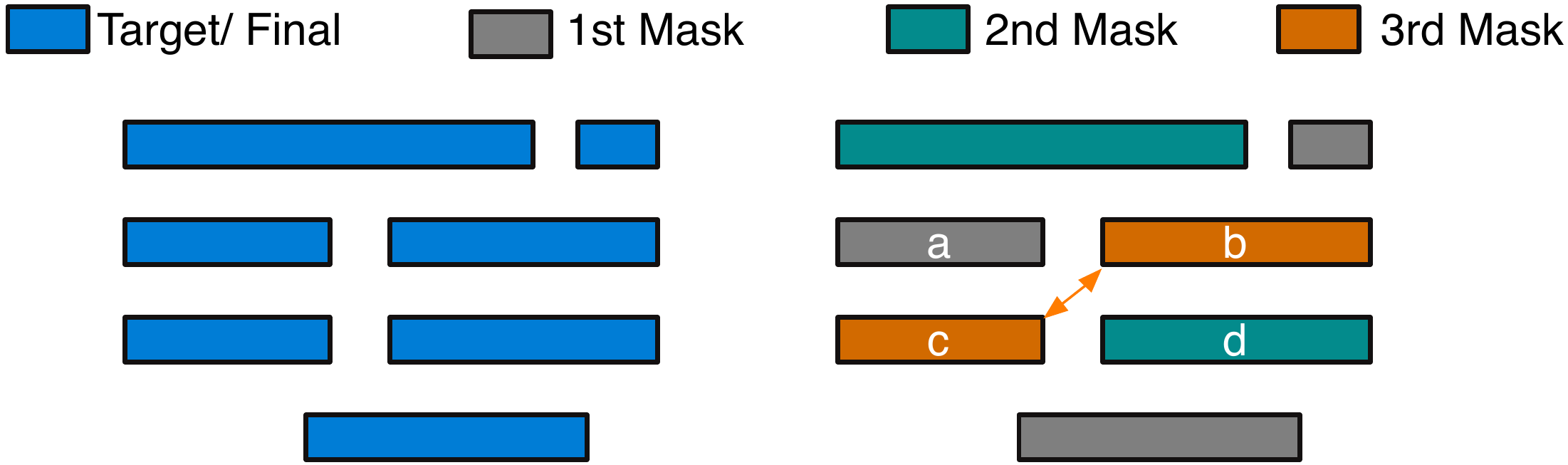}
  \\ (a) \hspace{4.4cm} (b)
  \caption{Process of LELELE type triple patterning lithography~(a) Target features;~(b) Layout decomposition with one conflict introduced.}
  \label{fig:tplec_le3}
\end{figure}

One conventional process of TPL, called LELELE, is with the same principle of litho-etch-litho-etch (LELE) type double patterning lithography (DPL).
Here ``L'' and ``E'' represent one lithography process and one etch process, respectively.
Although LELELE process has been widely studied by industry and academia, there are twofold issues derived.
On one side, even with stitch insertion, there are some native conflicts in LELELE, like 4-clique conflict \cite{TPL_ICCAD2011_Yu}.
For example, Fig. \ref{fig:tplec_le3} illustrates a 4-clique conflict among features $a$, $b$, $c$, and $d$.
No matter how to assign the colors, there is at least one conflict.
Since this 4-clique structure is common in advanced standard cell design, LELELE type TPL still suffers from the native conflict problem.
On the other side, compared with LELE type double patterning, there are more serious overlapping problem in LELELE \cite{TPL_SPIE08_Ausschnitt}.

\begin{figure}[tb]
  \centering
  \includegraphics[width=0.80\textwidth]{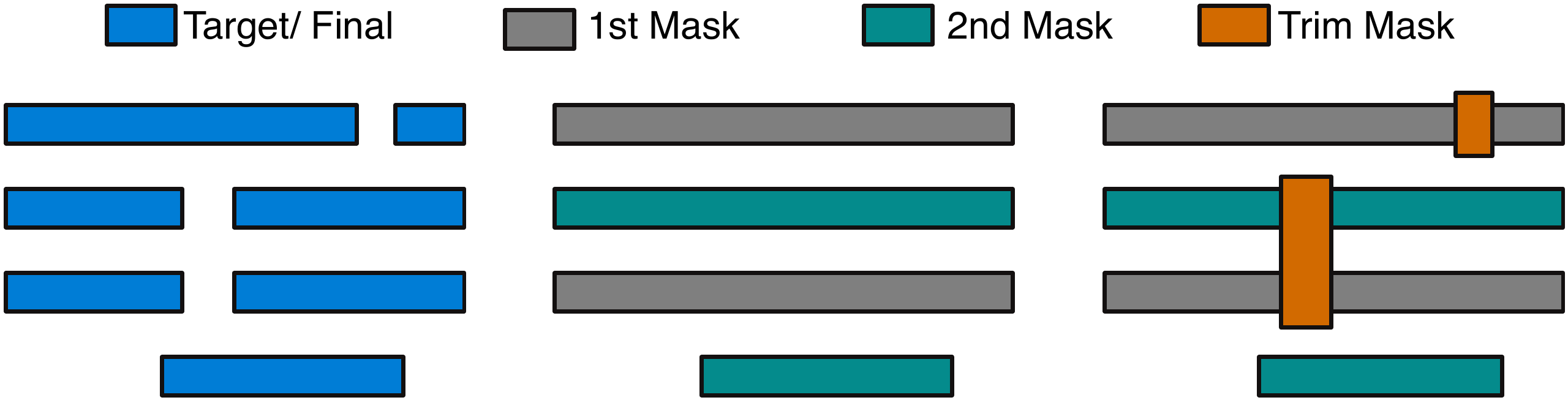}
  \\ (a) \hspace{3.8cm} (b) \hspace{3.8cm} (c)
  \caption{Process of LELE-EC type triple patterning lithography~(a) Target features;~(b) First and second mask patterns;~(c) Trim mask, and final decomposition without conflict.}
  \label{fig:tplec}
\end{figure}

To overcome all these limitations from LELELE, recently Lin \cite{LITH_ISPD2012_Lin} proposed a new TPL manufacturing process, called LELE-end-cutting (LELE-EC).
As a TPL, this new manufacturing process contains three mask steps, namely first mask, second mask, and \textit{trim} mask.
Fig. \ref{fig:tplec} illustrates an example of the LELE-EC process.
To generate target features in Fig. \ref{fig:tplec}(a), the first and second masks are used for pitch splitting,
which is similar to LELE type DPL process.
These two masks are shown in Fig. \ref{fig:tplec}(b).
Finally, a trim mask is applied to trim out the desired region as in Fig. \ref{fig:tplec}(c).
In other words, the trim mask is used to generate some end-cuts to further split feature patterns.
Although the target features are not LELELE-friendly, they are LELE-EC process friendly that with LELE-EC the features can be decomposed without any conflict.
In addition, if all cuts are properly designed or distributed, LELE-EC can introduce better printability then conventional LELELE process \cite{LITH_ISPD2012_Lin}.

\begin{figure}[tb]
  \centering
  \includegraphics[width=0.80\textwidth]{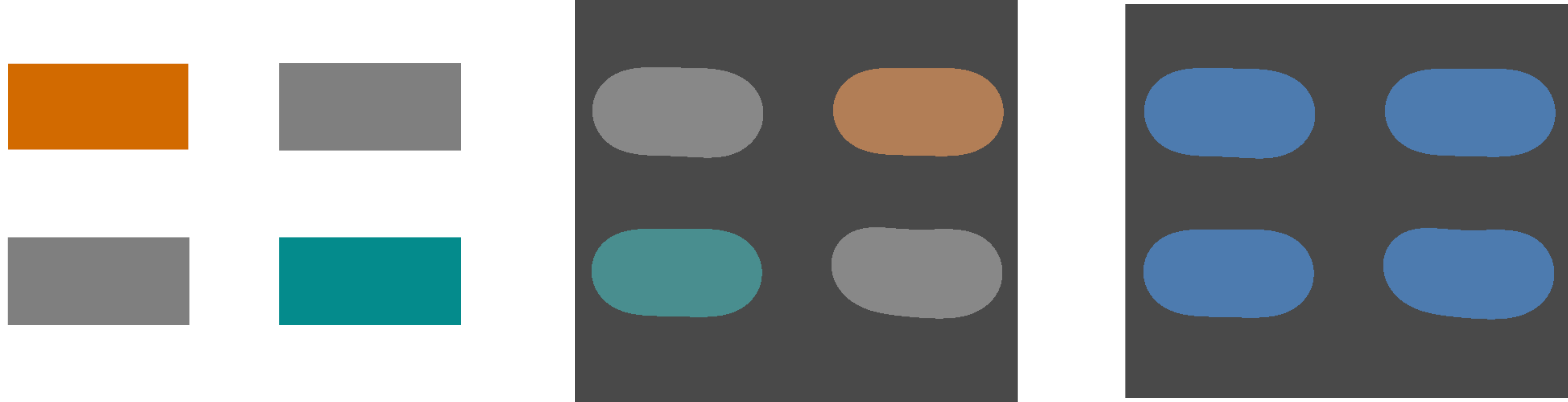}
  \\ (a) \hspace{3.8cm} (b) \hspace{3.8cm} (c)
  \caption{LELELE process example.
  (a) Decomposed result;
  (b) Simulated images for different masks;
  (c) Combined simulated image as the final printed patterns.
  }
  \label{fig:tplec_le3_img}
\end{figure}

\begin{figure}[tb]
  \centering
  \includegraphics[width=0.80\textwidth]{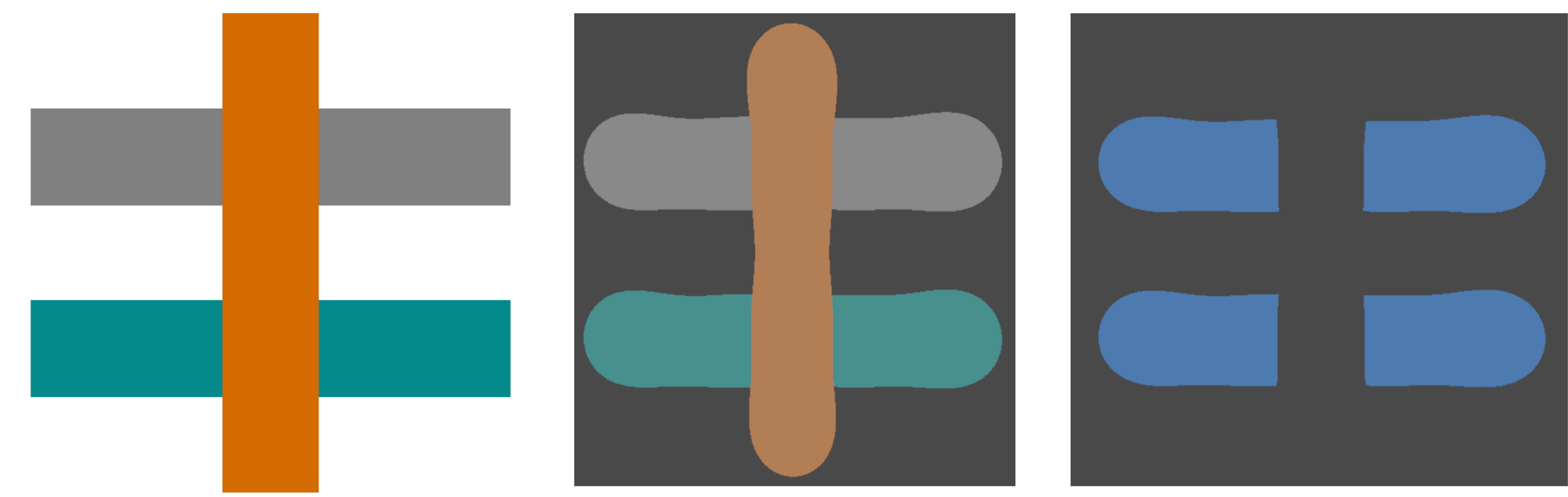}
  \\ (a) \hspace{3.8cm} (b) \hspace{3.8cm} (c)
  \caption{LELE-EC process example.
  (a) Decomposed result;
  (b) Simulated images for different masks, where orange pattern is trim mask;
  (c) Combined simulated image as the final printed patterns.
  }
  \label{fig:tplec_ec_img}
\end{figure}

For a design with four short features, Fig. \ref{fig:tplec_le3_img} and Fig. \ref{fig:tplec_ec_img} present its simulated images through LELELE and LELE-EC processes, respectively.
The lithography simulations are computed based on the partially coherent imaging system, where the 193nm illumination source is modeled as a kernel matrix given by [\citenum{OPC_ICCAD2013_Banerjee}].
To model the photoresist effect with the exposed light intensity, we use the constant threshold model with threshold $0.225$.
We can make several observations from these simulated images.
First, there are some round-offs around the line ends (see Fig. \ref{fig:tplec_le3_img} (c)).
Second, to reduce the round-off issues, as illustrated in Fig. \ref{fig:tplec_ec_img} (b), in LELE-EC process the short lines can be merged into longer lines,
then the trim mask is used to cut off some spaces.
It shall be noted that there might be some corner roundings due to the edge shorting of trim mask patterns.
However, since the line shortening or rounding is a strong function of the line width \cite{LITH_Book08_Mack},
and we observe that usually trim mask patterns can be much longer than line-end width,
thus we assume the rounding caused by the trim mask is insignificant.
This assumption is demonstrated as in Fig. \ref{fig:tplec_ec_img} (c).

Many research have been carried out to solve the corresponding design problems for LELELE type TPL.
The layout decomposition problem has been well studied
\cite{TPL_SPIE08_Cork,TPL_ICCAD2011_Yu,TPL_DAC2012_Fang,TPL_ICCAD2012_Tian,TPL_DAC2013_Kuang,TPL_ICCAD2013_Yu,TPL_ICCAD2013_Zhang,MPL_DAC2014_Yu}.
In addition, the related constraints have been considered in early physical design stages, like routing \cite{DFM_DAC2012_Ma,DFM_ICCAD2012_Lin},
standard cell design \cite{TPL_ICCAD2013_Tian,DFM_ICCAD2013_Yu}, and detailed placement \cite{DFM_ICCAD2013_Yu}.
However, only few attempts have been made to address the LELE-EC layout decomposition problem.
It shall be noted that although trim mask can bring about better printability, it does introduce more design challenge, especially in layout decomposition stage.

In this paper, we propose a comprehensive study for LELE-EC layout decomposition.
Given a layout which is specified by features in polygonal shapes, we extract the geometrical relationships and construct the conflict graphs.
Furthermore, the compatibility of all end-cuts candidates are also modeled in the conflict graphs.
Based on the conflict graphs, integer linear programming (ILP) is formulated to assign each vertex into one layer.
Our goal in the layout decomposition is to minimize the conflict number, and at the same time minimize the overlapping errors.

The rest of the paper is organized as follows.
Section \ref{sec:tplec_prelim} provides some preliminaries, and discusses the problem formulation.
Section \ref{sec:tplec_overview} provides the overall flow of our layout decomposer.
Section \ref{sec:tplec_endcut} explains the details of end-cut candidate generation.
Section \ref{sec:tplec_algo} presents a set of algorithms to solve the layout decomposition problem.
Section \ref{sec:tplec_speedup} discusses several speed-up techniques.
Section \ref{sec:tplec_result} presents the experimental results, followed by conclusions in Section \ref{sec:tplec_conclu}.

\section{Preliminary and Problem Formulation}
\label{sec:tplec_prelim}

\subsection{Layout Graph}

\begin{figure}[htb]
  \centering
  \includegraphics[width=0.65\textwidth]{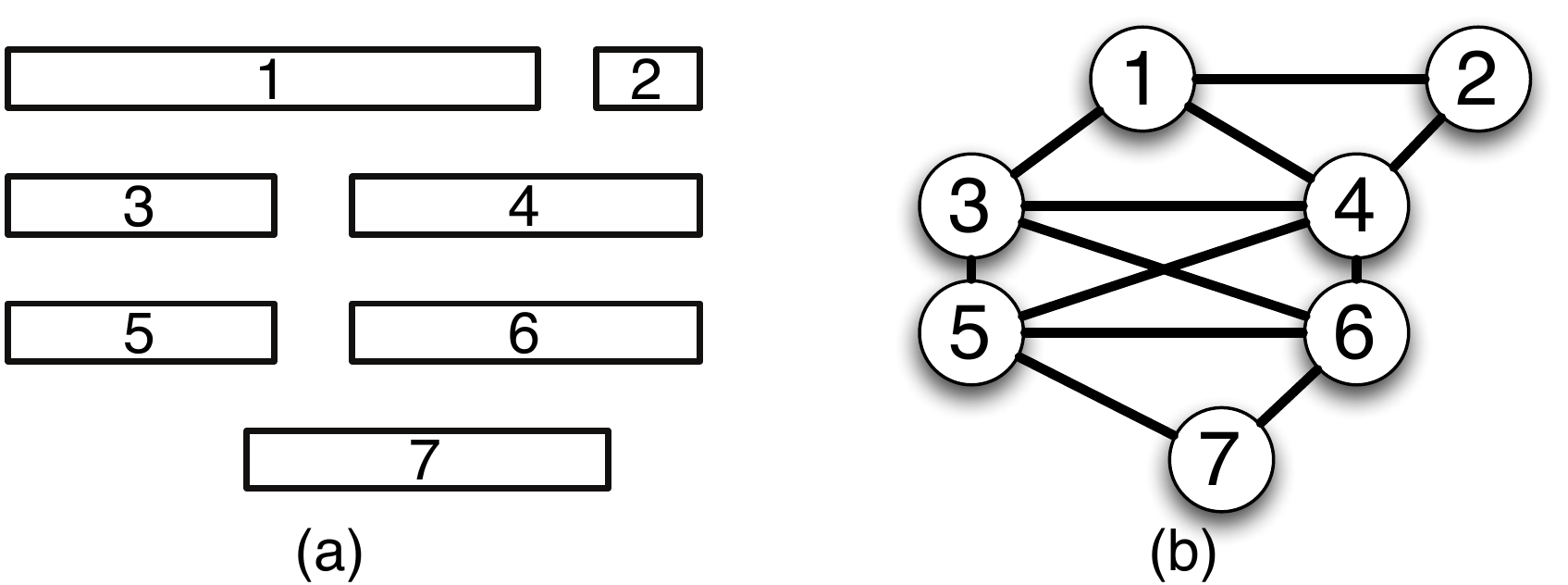}
  \caption{Layout graph construction.
  (a) Input layout;
  (b) Layout graph with conflict edges;
  }
  \label{fig:tplec_lg}
\end{figure}

Given a layout which is specified by features in polygonal shapes, layout graph \cite{TPL_ICCAD2011_Yu} is constructed.
As shown in Fig. \ref{fig:tplec_lg},
The layout graph is an undirected graph with a set of vertices $V$ and a set of conflict edges $CE$.
Each vertex in $V$ represents one input feature.
There is an edge in $CE$ if and only if the two features are within minimum coloring distance $dis_m$ of each other.
In other words, each edge in $CE$ is a conflict candidate.
Fig. \ref{fig:tplec_lg}(a) shows one input layout, and the corresponding layout graph is in Fig. \ref{fig:tplec_lg}(b).
Here the vertex set $V=\{1,2,3,4,5,6,7\}$,
while the conflict edge set $CE=\{(1,2),(1,3),(1,4),(2,4),(3,4),(3,5),(3,6),(4,5),(4,6),(5,6),(5,7),(6,7)\}$.
For each edge (conflict candidate), we check whether there is an end-cut candidate.
For each end-cut candidate $i-j$, if it is applied, then features $i$ and $j$ will be merged into one feature.
By this way the corresponding conflict edge can be removed.
If stitch is considered in layout decomposition, some vertices in layout graph can be split into several segments.
The segments in one layout graph vertex are connected through \textit{stitch edges}.
All these stitch edges are included in a set, called $SE$.
Please refer to [\citenum{DPL_ICCAD08_Kahng}] for the details of stitch candidate generation.

\subsection{End-Cut Graph}
\label{sec:ec_graph}

\begin{figure}[htb]
  \centering
  \includegraphics[width=0.95\textwidth]{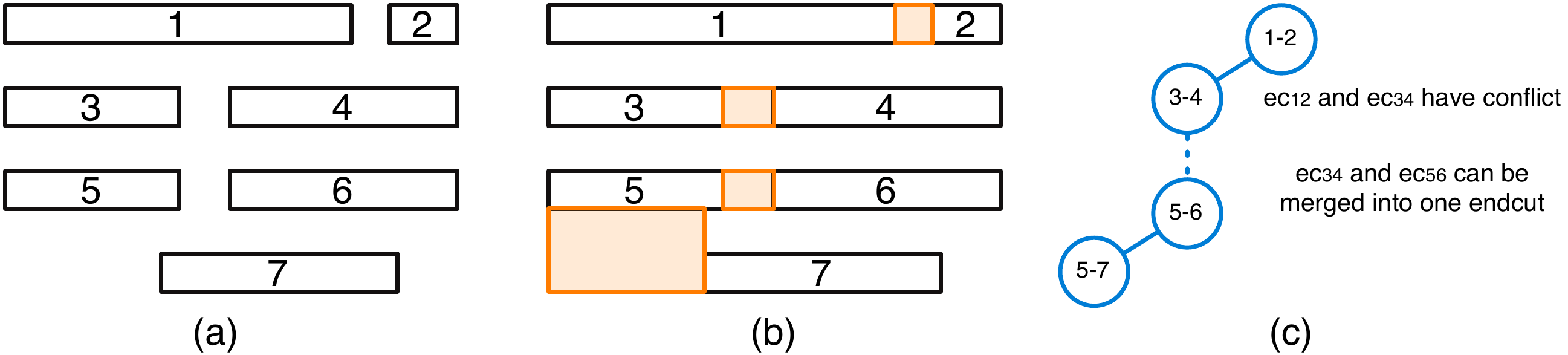}
  \caption{End-cut graph construction.
  (a) Input layout;
  (b) Generated end-cut candidates.
  (c) End-cut graph.
  }
  \label{fig:tplec_ecgraph}
\end{figure}

Since all the end-cuts are manufactured through one single exposure process, they should be distributed far away from each other.
That is, two end-cuts have conflict if they are within minimum end-cut distance $dis_c$ of each other.
Note that these conflict relationships among end-cuts are not available in layout graph, therefore we construct \textit{end-cut graph} to store the relationships.
Fig. \ref{fig:tplec_ecgraph}(a) gives an input layout example,
with all end-cut candidates pointed out in Fig. \ref{fig:tplec_ecgraph} (b);
and the corresponding end-cut graph is shown in Fig. \ref{fig:tplec_ecgraph} (c).
Each vertex in the end-cut graph represents one end-cut.
There is an solid edge if and only if the two end-cuts conflict to each other.
There is an dash edge if and only if they are close to each other, and they can be merged into one larger end-cut. 

\subsection{Problem Formulation}

Here we give the problem formulation of layout decomposition for triple patterning with End-Cutting (LELE-EC).

\begin{myproblem}[LELE-EC Layout Decomposition]
Given a layout which is specified by features in polygonal shapes, the layout graph and the end-cut graph are constructed.
The LELE-EC layout decomposition assigns all vertices in layout graph into one of two colors, and select a set of end-cuts in end-cut graph.
The objectives is to minimize the number of conflict and/or stitch.
\end{myproblem}

With the end-cut candidates generated, the LELE-EC layout decomposition is more complicated since more constraints are derived.
Even there is no end-cut candidate, LELE-EC layout decomposition is similar to the LELE type DPL layout decomposition.
Sun et al. in Ref. [\citenum{DFM_ASPDAC2011_Sun}] showed that LELE layout decomposition with minimum conflict and minimum stitch is NP-hard,
thus it is not hard to see that LELE-EC layout decomposition is NP-hard as well.
NP-hard problem is a set of computational search problems that are difficult to solve
\cite{book90Algorithm}.
A problem is NP-hard if solving it in polynomial time would make it possible to solve all problems in class NP in polynomial time. 
In other words, unless ``P=NP'', there is no polynomial time algorithm to solve an NP-hard problem optimally.

\section{Overall Flow}
\label{sec:tplec_overview}

\begin{figure}[htb]
  \centering
  \includegraphics[width=0.75\textwidth]{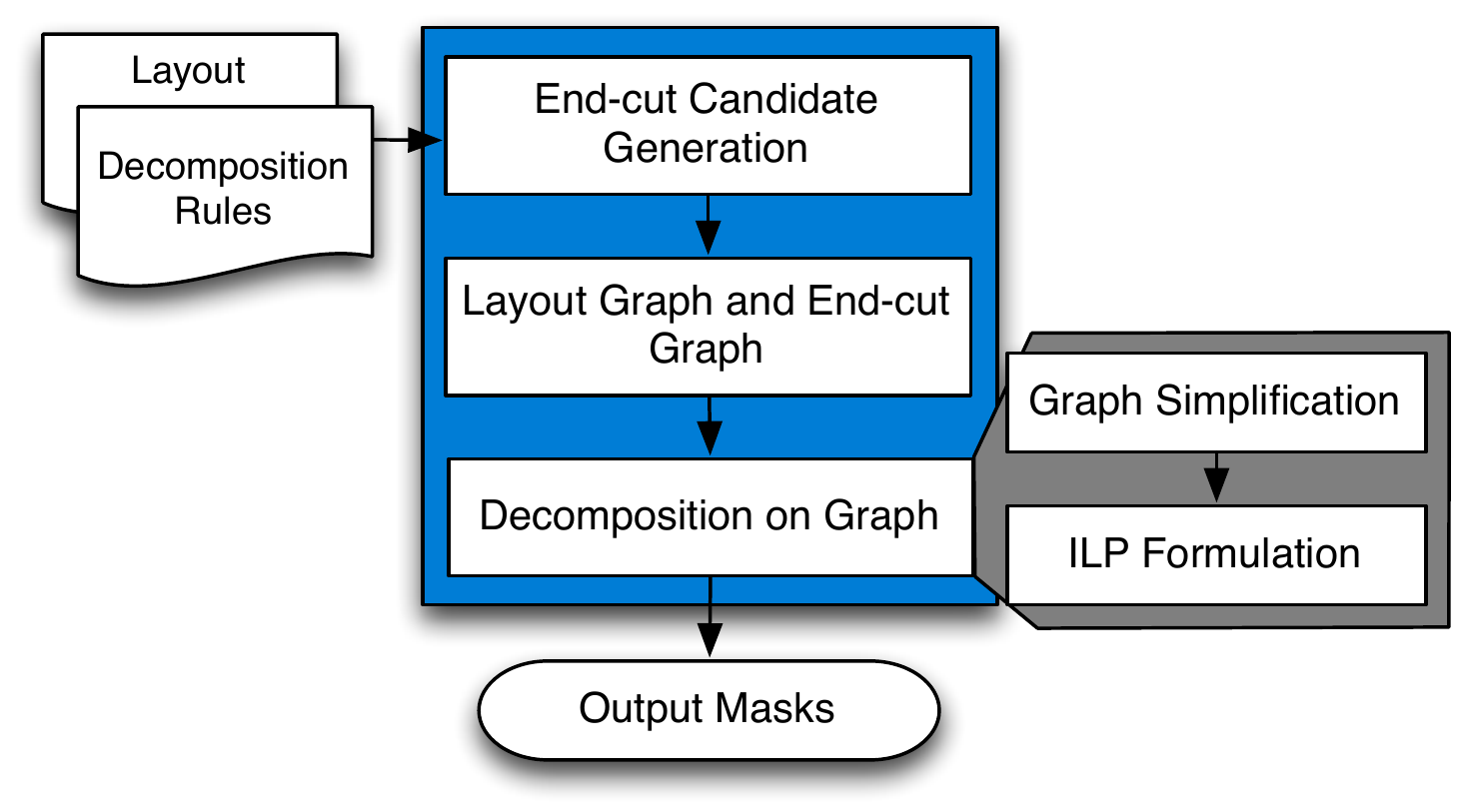}
  \caption{Overall flow of our layout decomposer.}
  \label{fig:tplec_flow}
\end{figure}

The overall flow of our layout decomposer is illustrated in Fig. \ref{fig:tplec_flow}.
First we generate all end-cut candidates to find out all possible end-cuts.
Then we construct the layout graph and the end-cut graph to transform the original geometric problem into a graph problem,
and thus the LELE-EC layout decomposition can be modeled as a coloring problem in layout graph and the end-cut selection problem in end-cut graph.
Both the coloring problem and the end-cut selection problem can be solved through one ILP formulation.
Since the ILP formulation may suffer from runtime overhead problem, we propose a set of graph simplification techniques.
Besides, to further reduce the problem size, some end-cut candidates are pre-selected before ILP formulation.
All the steps in the flow are detailed in the following sections.

\section{End-Cut Candidate Generation}
\label{sec:tplec_endcut}

In this section we will explain the details of our algorithm to generate all end-cut candidates.
An end-cut candidate is generated between two conflicting polygonal shapes.
It should be stressed that compared with the end-cut generation in Ref. [\citenum{TPLEC_SPIE2013_Yu}], our methodology has the following two differences.

\begin{figure}[htb]
  \centering
  \includegraphics[width=0.54\textwidth]{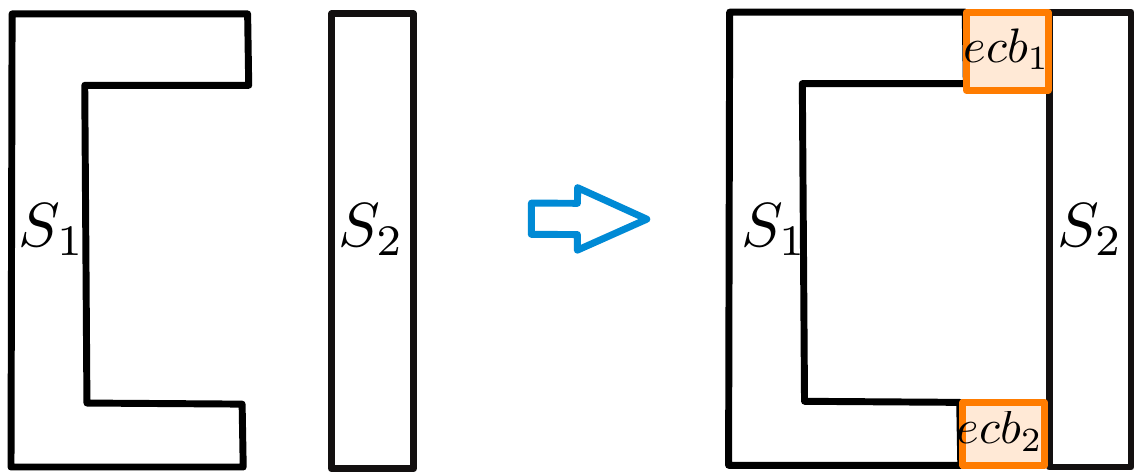}
  \\ (a) \hspace{4.4cm} (b)
  \caption{An end-cut can have multiple end-cut boxes.}
  \label{fig:multiplebox}
\end{figure}

\begin{figure}[htb]
  \centering
  \includegraphics[width=0.84\textwidth]{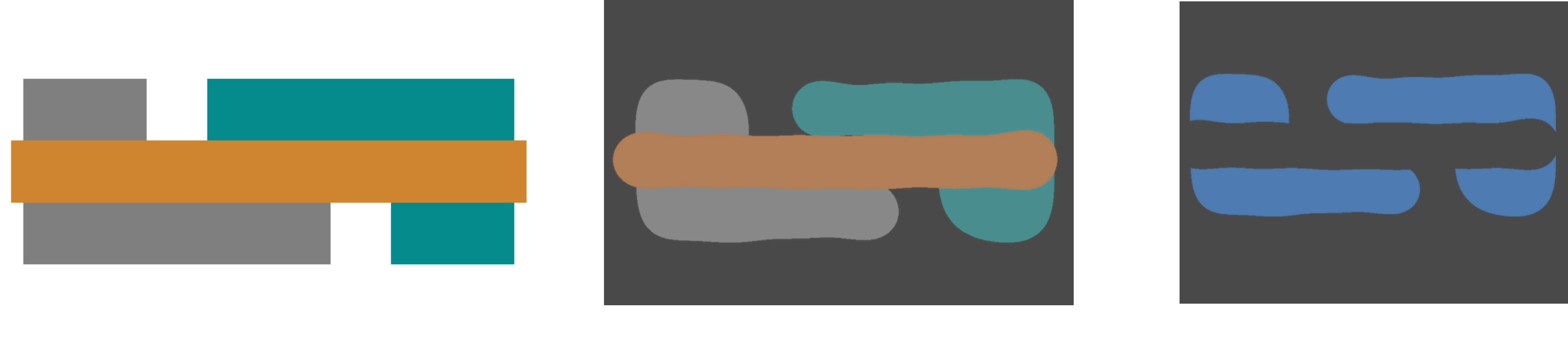}
  \\ (a) \hspace{4.0cm} (b) \hspace{4.0cm} (c)
  \caption{
  (a) Decomposition example where cuts are along long edges;
  (b) Simulated images for different masks;
  (c) Combined simulated image with some hotspots.
  }
  \label{fig:bad2_sim}
\end{figure}

\begin{figure}[htb]
  \centering
  \includegraphics[width=0.84\textwidth]{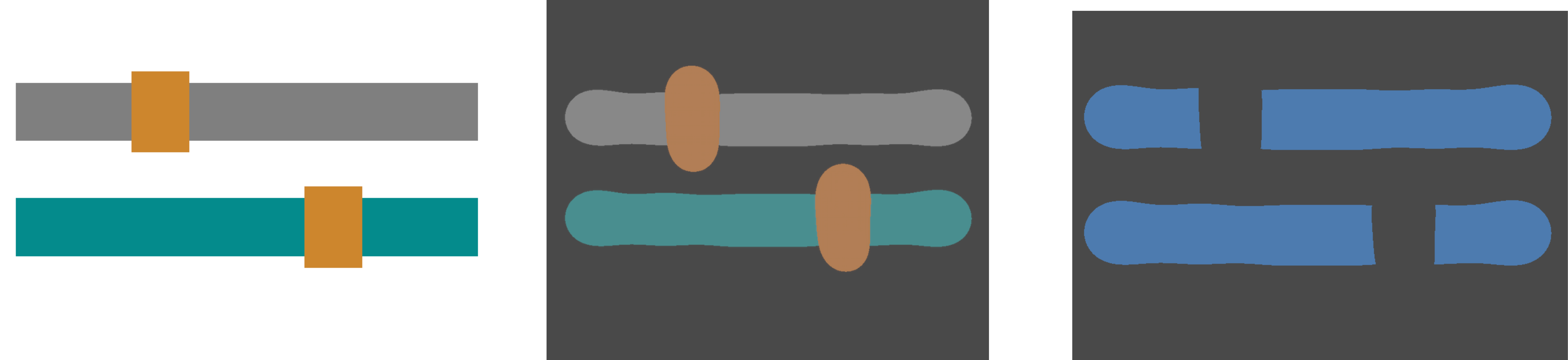}
  \\ (a) \hspace{4.0cm} (b) \hspace{4.0cm} (c)
  \caption{
  (a) Decomposition example where cuts are between line ends;
  (b) Simulated images for different masks;
  (c) Combined simulated image with good printability.
  }
  \label{fig:good2_sim}
\end{figure}

\begin{itemize}
  \item
An end-cut can be a collection of multiple end-cut boxes depending on the corresponding shapes.
For instance, two end-cut boxes ($ecb_{1}$ and $ecb_{2}$) need to be generated between shapes $S_{1}$ and $S_{2}$ as shown in Fig \ref{fig:multiplebox}.
We propose a shape-edge dependent algorithm to generate the end-cuts with multiple end-cut boxes.
  \item
We consider the overlapping and variations caused by end-cuts.
Two lithography simulations are illustrated in Fig. \ref{fig:bad2_sim} and Fig. \ref{fig:good2_sim}, respectively.
In Fig. \ref{fig:bad2_sim} we find some bad patterns or hotspots, due to the cuts between two long edges.
In Fig. \ref{fig:good2_sim} we can see that the final patterns are in better shape. 
Therefore, to reduce the manufacturing hotspot from trim mask, during end-cut candidate generation we avoid the cuts along two long edges.
\end{itemize}

Algorithm \ref{alg:endcutgeneration} presents the key steps of generating end-cut between two polygonal shapes $S_{1}$ and $S_{2}$.
A polygonal shape consists of multiple edges. For each of the shape-edge-pair, one taken from $S_{1}$ and another from $S_{2}$,
the possibility of generation of end-cut box ($ecBox$) is explored and we store all such end-cut boxes in $ecBoxSet$ (Lines $2$-$9$).
The function `generateEndCutBox($se_{1}$, $se_{2}$)' generates an end-cut box $ecBox$ between the shape-edges $se_{1}$ and $se_{2}$. 

\begin{figure}[htb]
  \centering
  \includegraphics[width=0.94\textwidth]{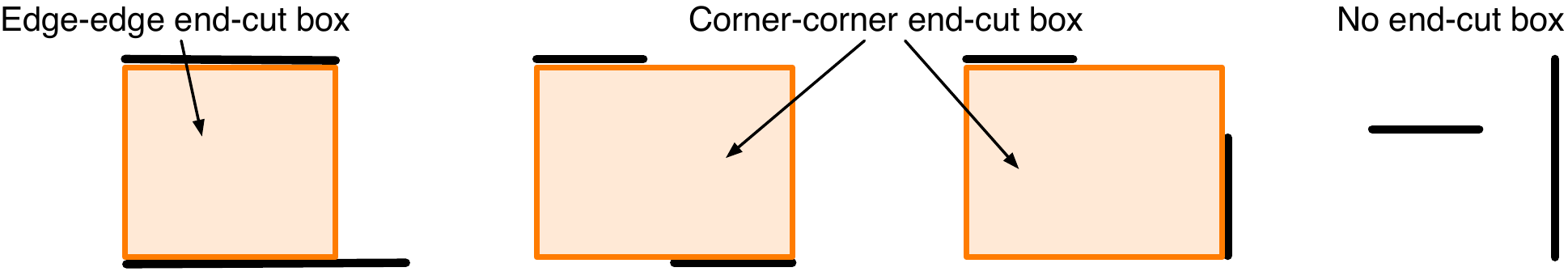}
  \\ \hspace{1.4cm} (a) \hspace{3.4cm} (b) \hspace{3.4cm} (c) \hspace{3.4cm} (d)
  \caption{End-cut box generation between any two shape-edges}
  \label{fig:endcutboxgeneration}
\end{figure}

Fig. \ref{fig:endcutboxgeneration} shows how end-cut boxes are generated between two shape-edges under different situations. In Fig. \ref{fig:endcutboxgeneration}(a), the end-cut box is between two shape-edges which are oriented in same direction and overlap in its x-coordinates. This type of end-cut box is called as edge-edge end-cut box. For Fig. \ref{fig:endcutboxgeneration}(b) the shape edges are in same direction but they do not overlap, and in Fig. \ref{fig:endcutboxgeneration}(c), the shape-edges are oriented in different directions. The end-cut boxes generated in these two cases are called corner-corner end-cut boxes. No end-cut box is generated in case of Fig. \ref{fig:endcutboxgeneration}(d). In addition, end-cut boxes are not generated for the following cases: (i) the end-cut box is overlapping with any existing polygonal shape in the layout, (ii) the height ($h$) or width ($w$) of the box is not within some specified range, \textit{i.e.}, when $h$, $w$ do not obey the following constraints $h_{low} \le h \le h_{high}$ and $w_{low} \le w \le w_{high}$.

\begin{algorithm}[htb]
\caption{Shape-edge dependent end-cut generation algorithm between two shapes $S_{1}$ and $S_{2}$}\label{alg:endcutgeneration}
\begin{algorithmic}[1]
\State \textbf{Procedure} generateEndCut ($S_{1}$, $S_{2}$);
\ForAll{$se_{1}$ $\in$ edges($S_{1}$)}
        \ForAll{$se_{2}$ $\in$ edges($S_{2}$)}
                \State $ecBox$ = generateEndCutBox ($se_{1}$, $se_{2}$);
                \If{$ecBox$ $\ne$ NULL}
                        \State Store $ecBox$ in $ecBoxSet$;
                \EndIf
        \EndFor
\EndFor
\State Divide $ecBoxSet$ into independent components ($IC$);
\If{$|ecBoxSet| = |V|$}
        \State Print all boxes;
        \State \textbf{return} true;
\EndIf
\ForAll{$ic \in IC$}
        \State Remove corner-corner end-cut boxes overlapping with edge-edge end-cut box;
        \If{$\exists$ set of $type_{2}$ overlaps}
                \State Generate minimum area box;
        \Else
                \State Generate all end-cut boxes
        \EndIf
\EndFor
\State \textbf{return} true;
\State \textbf{end Procedure}
\end{algorithmic}
\end{algorithm}

\begin{figure}[htb]
  \centering
  \includegraphics[width=0.8\textwidth]{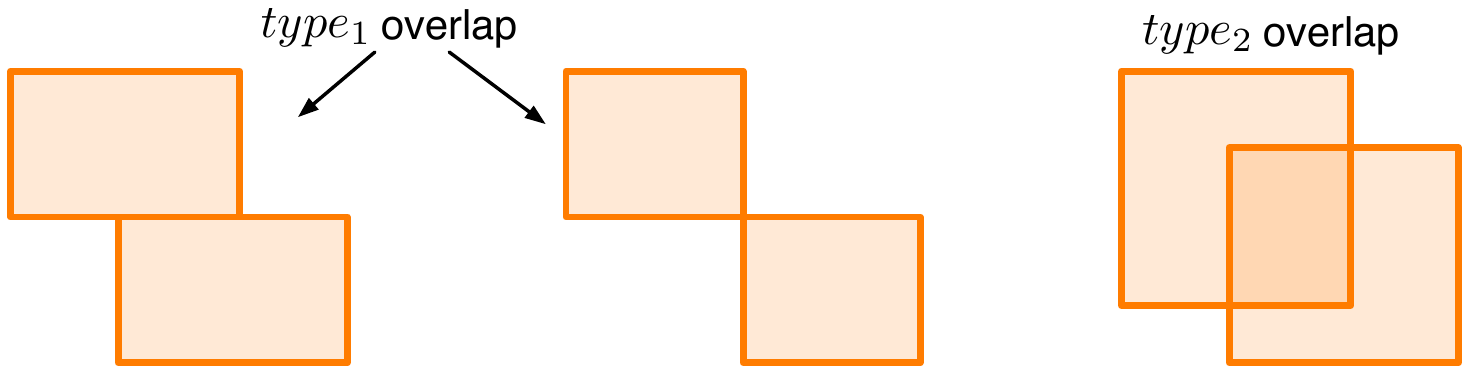}
  \\ \hspace{0.4cm} (a) \hspace{3.8cm} (b) \hspace{3.8cm} (c)
  \caption{Types of overlaps between end-cut boxes}
  \label{fig:overlaptype}
\end{figure}

Then all generated end-cut boxes between two shapes are divided into independent components $IC$ (Line $10$) based on finding connected components of a graph $G=(V,E)$ with $V=\{v_{i}\}=$ set of all end-cut boxes and ($v_{i},v_{j}) \in E$, if $v_{i}$ overlaps with $v_{j}$. The overlap between two end-cut boxes is classified into $type_{1}$ and $type_{2}$ overlaps. When two boxes overlap only in an edge or in a point but not in space, we call this $type_{1}$ overlap, where as the overlap in space is termed as $type_{2}$ overlap as shown in Fig. \ref{fig:overlaptype}. Each of the $ic \in IC$ may contain multiple end-cut boxes. If the total number of end-cut-boxes ($|V|$) is equal to $|IC|$, that implies there is no overlap between the end-cut boxes and we generate all of them (Lines $11$-$14$). 

\begin{figure}[htb]
  \centering
  \includegraphics[width=0.84\textwidth]{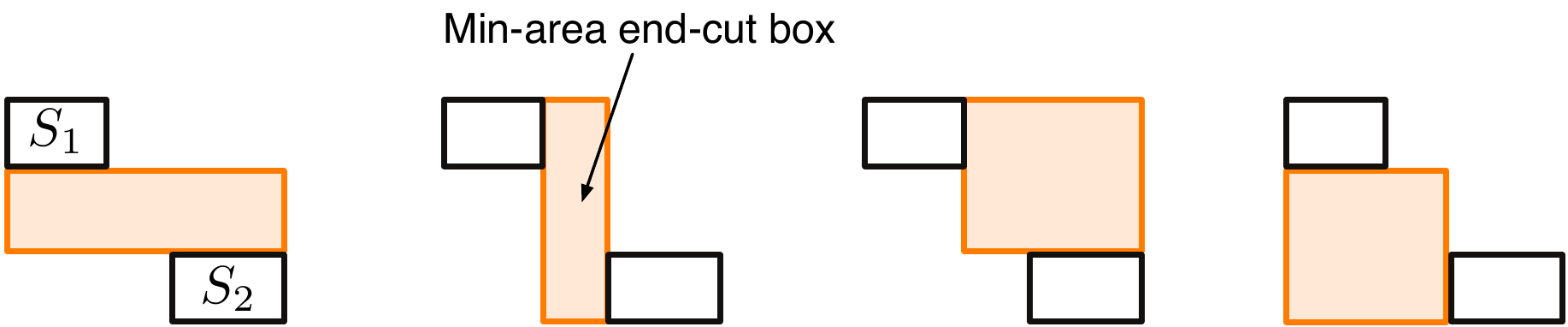} 
  \\ \hspace{1.4cm} (a) \hspace{3.0cm} (b) \hspace{3.0cm} (c) \hspace{3.0cm} (d)
  \caption{Minimum area end-cut box is chosen for $type_{2}$ overlaps}
  \label{fig:type2overlaps}
\end{figure}

For multiple boxes in each $ic$, if there is an overlap between corner-corner and edge-edge end-cut boxes, the corner-corner end-cut box is removed (Line $16$).
After doing this, either there will be a set of $type_{1}$ overlaps or a set of $type_{2}$ overlaps in each $ic$.
In case of $type_{2}$ overlaps, the end-cut box with minimum area is chosen as shown in Fig. \ref{fig:type2overlaps}.
For $type_{1}$ overlaps in each $ic$, all end-cut boxes are generated (Line $20$).

\section{ILP Formulations}
\label{sec:tplec_algo}

After the construction of layout graph and end-cut graph, LELE-EC layout decomposition problem can be transferred into a coloring problem on layout graph and a selection problem on end-cut graph.
At the first glance the coloring problem is similar to that in LELE layout decomposition.
However, since the conflict graph can not be guaranteed to be planar, the face graph based methodology \cite{DPL_ISPD2010_Xu} cannot be applied here.
Therefore, we formulate integer linear programming (ILP) to solve both coloring problem and selection problem simultaneously.
For convenience, some notations in the ILP formulation are listed in Table \ref{tab:tplec_notation}.

\begin{table}[hbt]
\centering
\caption{Notations in LELE-EC Layout Decomposition}
\label{tab:tplec_notation}
\begin{tabular}{|c|c|}
	\hline
  \hline  $CE$		    & set of conflict edges\\
	\hline  $EE$		    & set of end-cut conflict edges\\
  \hline  $SE$		    & set of stitch edges.\\
	\hline  $n$         & number of input layout features. \\
  \hline  $r_i$		    & the $i_{th}$ layout feature\\
	\hline  $x_i$		    & variable denoting the coloring of $r_i$\\
	\hline  $ec_{ij}$   & 0-1 variable, $ec_{ij}=1$ when a end-cut between $r_i$ and $r_j$\\
  \hline  $c_{ij}$	  & 0-1 variable, $c_{ij}=1$ when a conflict between $r_i$ and $r_j$\\
	\hline  $s_{ij}$	  & 0-1 variable, $s_{ij}=1$ when a stitch between $r_i$ and $r_j$\\
	\hline\hline
\end{tabular}
\end{table}

\subsection{ILP Formulation w/o. Stitch}
In this subsection we discuss the ILP formulation when no stitch candidate is generated in layout graph.
Given a set of input layout features $\{r_1, \dots, r_n\}$, we construct layout graph, and end-cut graph.
Every conflict edge $e_{ij}$ is in $CE$, while each end-cut candidate $ec_{ij}$ is in $SE$.
$x_i$ is a binary variable representing the color of $r_i$.
$c_{ij}$ is a binary variable for conflict edge $e_{ij} \in CE$.
To minimize the conflict number, our objective function to minimize $\sum_{e_{ij} \in CE} c_{ij}$.

To evaluate the conflict number, we provide the following constraints:
\begin{equation}
  \label{cnst:c}
	\left\{
	\begin{array}{cl}
		x_i + x_j \le 1 + c_{ij} + ec_{ij}            & \textrm{if}\ \exists     \ ec_{ij} \in EE \\
    (1-x_i) + (1-x_j) \le 1 + c_{ij} + ec_{ij}    & \textrm{if}\ \exists     \ ec_{ij} \in EE \\
		x_i + x_j \le 1 + c_{ij}                      & \textrm{if}  \not\exists \ ec_{ij} \in EE \\
    (1-x_i) + (1-x_j) \le 1 + c_{ij}              & \textrm{if}  \not\exists \ ec_{ij} \in EE 
	\end{array}
	\right.
\end{equation}
Here $ec_{ij}$ is a binary variable for end-cut candidate.
If there is no end-cut candidate between adjacent features $r_i$ and $r_j$, if $x_i \ne x_j$ then one conflict would be reported ($c_{ij} = 1$).
Otherwise, we will try to enable the end-cut candidate $ec_{ij}$ first.
If the end-cut candidate $ec_{ij}$ cannot be applied ($ec_{ij} = 0$), then one conflict will be also reported.

If end-cuts $ec_{ij}$ and $ec_{pq}$ are conflict with each other, at most one of them will be applied.
To enable this, we introduce the following constraint.
\begin{equation}
  ec_{ij} + ec_{pq} \le 1,   \ \ \forall e_{ijpq} \in EE  \label{cnst:ec1} 
\end{equation}

To forbid useless end-cut, we introduce the following constraints.
That is, if features $x_{i}$ and $x_{j}$ are in different colors, $ec_{ij} = 0$.
\begin{equation}
  \label{cnst:ec2}
  \left\{
  \begin{array}{cc}
  ec_{ij} + x_i - x_j \le 1                     & \forall e_{ij} \in CE    \\
  ec_{ij} + x_j - x_i \le 1                     & \forall e_{ij} \in CE    
  \end{array}
  \right.
\end{equation}

Therefore, without the stitch candidate, the LELE-EC layout decomposition can be formulated as shown in Eq. (\ref{eq:ec_ilp1}).
\begin{align}
  \textrm{min}      &  \sum_{e_{ij} \in CE} c_{ij}                                    \label{eq:ec_ilp1}\\
  \textrm{s.t.} \ \ & (\ref{cnst:c}), (\ref{cnst:ec1}), (\ref{cnst:ec2})              \notag
\end{align}

\subsection{ILP Formulation w. Stitch}
If the stitch insertion is considered, the ILP formulation is as in Eq. (\ref{eq:ec_ilp3}).
Here the objective is to simultaneously minimize both the conflict number and the stitch number.
The parameter $\alpha$ is a user-defined parameter for assigning relative importance between the conflict number and the stitch number.
The constraints (\ref{eq:ec_ilp3}a) - (\ref{eq:ec_ilp3}b) are used to calculate the stitch number.

\begin{figure}[htb]
\centering
\begin{align}
  \textrm{min} & \sum_{e_{ij}\in CE} c_{ij} + \alpha \times \sum_{e_{ij} \in SE} s_{ij} &      \label{eq:ec_ilp3} \\
  \textrm{s.t.} \ \ &  x_i - x_j \le s_{ij}                          & \forall e_{ij} \in SE   \tag{\ref{eq:ec_ilp3}$a$} \\
                    &  x_j - x_i \le s_{ij}                          & \forall e_{ij} \in SE   \tag{\ref{eq:ec_ilp3}$b$} \\
                    & (\ref{cnst:c}), (\ref{cnst:ec1}), (\ref{cnst:ec2}) &\notag
\end{align}
\end{figure}

\section{Graph Simplification Techniques}
\label{sec:tplec_speedup}

ILP is a classical NP-hard problem, i.e., there is no polynomial time optimal algorithm to solve it \cite{book90Algorithm}.
Therefore, for large layout cases, solving ILP may suffer from long runtime penalty to achieve the results.
In this section, we provide a set of speed-up techniques.
Note that these techniques can keep optimality.
In other words, with these speed-up techniques, ILP formulation can achieve the same results comparing with those not applying speed-up.

\subsection{Independent Component Computation}

The first speed-up technique is called \textit{independent component computation}.
By breaking down the whole layout graph into several independent components, we partition the initial layout graph into several small ones.
Then each component can be resolved through ILP formulation independently.
At last, the overall solution can be taken as the union of all the components without affecting the global optimality.
Note that this is a well-known technique which has been applied in many previous studies ( e.g., Ref. [\citenum{DPL_ICCAD08_Kahng,DPL_ISPD09_Yuan,DPL_ASPDAC2010_Yang}]).

\subsection{Bridge Computation}

A bridge of a graph is an edge whose removal disconnects the graph into two components.
If the two components are independent, removing the bridge can divide the whole problem into two independent sub-problems.
We search all bridge edges in layout graph, then divide the whole layout graph through these bridges.
Note that all bridges can be found in $O(|V|+|E|)$, where $|V|$ is the vertex number, and $|E|$ is the edge number in the layout graph.

\subsection{End-Cut Pre-Selection}

Some end-cut candidates have no conflict end-cuts.
For the end-cut candidate $ec_{ij}$ that has no conflict end-cut, it would be pre-selected in final decomposition results.
That is, the features $r_i$ and $r_j$ are merged into one feature.
By this way, the problem size of ILP formulation can be further reduced.
End-cut pre-selection can be finished in linear time.

\section{Experimental Results}
\label{sec:tplec_result}

We implement our algorithms in C++ and test on an Intel Xeon 3.0GHz Linux machine with 32G RAM.
15 benchmark circuits from Ref. [\citenum{TPL_ICCAD2011_Yu}] are used.
GUROBI \cite{Gurobi} is chosen as the ILP solver.
The minimum coloring spacing $min_s$ is set as 120 for the first ten cases and as 100 for the last five cases,
as in Ref. [\citenum{TPL_ICCAD2011_Yu}] and Ref. [\citenum{TPL_DAC2012_Fang}].
The width threshold $w_{th}$, which is used in end-cut candidate generation, is set as $dis_m$.

\subsection{With Stitch or Without Stitch}

\begin{table*}[htb]
\centering
\renewcommand{\arraystretch}{0.9}
\caption{Comparison between w. stitch and w/o. stitch}
\label{tab:tplec_stitch}
\resizebox{16.4cm}{!} {
\begin{tabular}{|c|c|c |c|c|c|c||c|c|c|c||}
  \hline \hline
  \multirow{2}{*}{Circuit} &\multirow{2}{*}{Wire\#}  &\multirow{2}{*}{Comp\#} &\multicolumn{4}{c||}{ILP w/o. stitch} & \multicolumn{4}{c|}{ILP w. stitch}\\
  \cline{4-11} &        &         &conflict\# &stitch\#  & cost & CPU(s)        &conflict\#    &stitch\#      &cost  & CPU(s) \\
  \hline
 C1           &1109    &123        &1     &0    &1       &1.19       &1       &0     &1       &1.32   \\ 
 C2           &2216    &175        &1     &0    &1       &2.17       &1       &0     &1       &2.89   \\  
 C3           &2411    &270        &0     &0    &0       &3.11       &0       &0     &0       &3.62   \\  
 C4           &3262    &467        &0     &0    &0       &3.2        &0       &0     &0       &3.75   \\   
 C5           &5125    &507        &4     &0    &4       &8.72       &4       &0     &4       &8.81   \\  
 C6           &7933    &614        &1     &0    &1       &11.24      &1       &0     &1       &11.1   \\ 
 C7           &10189   &827        &2     &0    &2       &14.57      &2       &0     &2       &15.98  \\   
 C8           &14603   &1154       &2     &0    &2       &21.2       &2       &0     &2       &23.07  \\  
 C9           &14575   &2325       &23    &0    &23      &24.36      &12      &12    &13.2    &28.06  \\   
 C10          &21253   &1783       &7     &0    &7       &28.42      &7       &0     &7       &32.02  \\    
 S1           &4611    &272        &0     &0    &0       &6.23       &0       &0     &0       &7.04   \\   
 S2           &67696   &5116       &166   &0    &166     &179.05     &166     &1     &166.1   &218.37 \\    
 S3           &157455  &15176      &543   &0    &543     &506.55     &530     &13    &531.3   &563.65 \\    
 S4           &168319  &15354      &443   &0    &443     &464.84     &436     &7     &436.7   &494.4  \\   
 S5           &159952  &12626      &419   &0    &419     &464.11     &415     &6     &415.6   &514.56 \\    
 \hline
 avg.         &-       &-          &107.5 &0    &107.5   &115.9      &105.1   &2.6   &105.4   &128.6  \\  
 \hline
 ratio        &-       &-          &-	    &-    &\textbf{1.0}   &\textbf{1.0}
                                   &-     &-    &\textbf{0.98}  &\textbf{1.10}\\

 \hline \hline
\end{tabular}
}
\end{table*}

In the first experiment, we show the decomposition results of the ILP formulation.
Table \ref{tab:tplec_stitch} compares two ILP formulations ``\textbf{ILP w/o. stitch}'' and ``\textbf{ILP w. stitch}''.
Here ``ILP w/o. stitch'' is the ILP formulation based on the graph without stitch edges, while ``ILP w. stitch'' considers the stitch insertion in the ILP.
Note that all speed-up techniques are applied to both.
Columns ``Wire\#'' and ``Comp\#'' reports the total feature number, and the divided component number, respectively.
For each method we report the conflict number, stitch number, and computational time in seconds(``CPU(s)'').
``Cost'' is the cost function, which is set as conflict\# $+ 0.1 \times $ stitch\#.

From Table \ref{tab:tplec_stitch} we can see that
compared with ``ILP w/o. stitch'', when stitch candidates are considered in the ILP formulation, the cost can be reduced by 2\%, 
while the runtime is increased by 5\%.
It shall be noted that stitch insertion has been shown to be an effective method to reduce the cost for both LELE layout decomposition and LELELE layout decomposition.
However, we can see that for LELE-EC layout decomposition, stitch insertion is not that effective.
In addition, due to the overlap variation derived from stitch, stitch insertion for LELE-EC may not be an effective method.

\subsection{Effectiveness of Speed-up Techniques}

\begin{figure}[htb]
  \centering
  \includegraphics[width=0.84\textwidth]{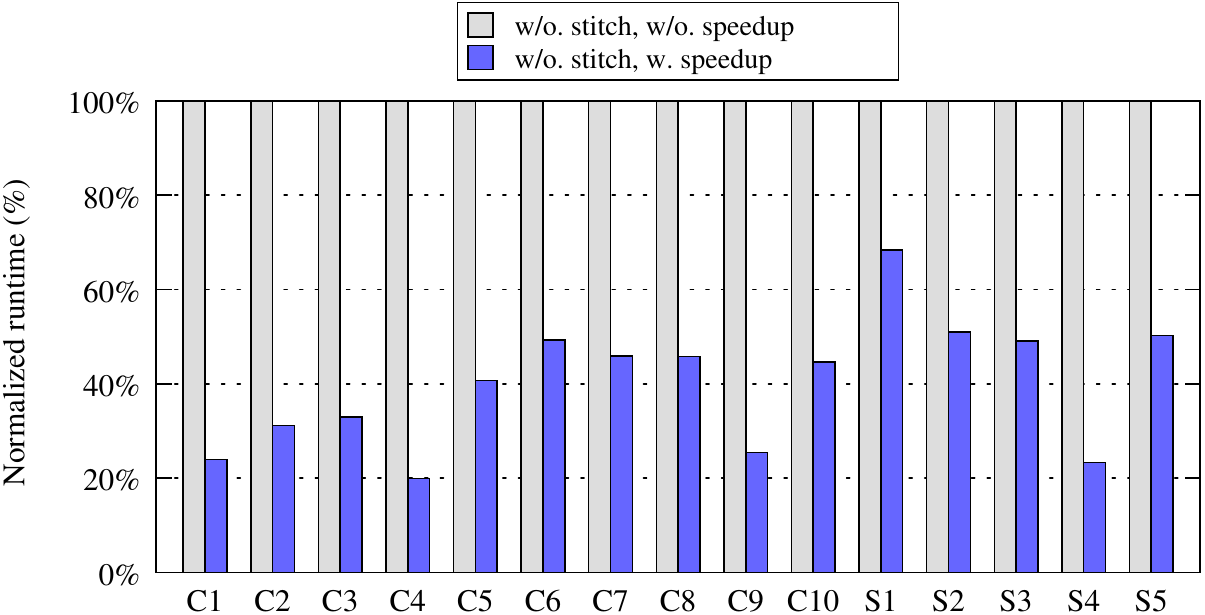}
  \caption{Effectiveness of Speed-up techniques when no stitch is introduced.}
  \label{fig:tplec_speedup_nostitch}
\end{figure}

In the second experiment, we analyze the effectiveness of the proposed speed-up techniques.
Fig. \ref{fig:tplec_speedup_nostitch} compares two ILP formulations ``\textbf{w/o. stitch w/o. speedup}'' and ``\textbf{w/o. stitch w. speedup}'',
where ``w/o. stitch w/o. speedup'' only applies independent component computation, while ``w. speedup'' involves all three speed-up techniques.
Besides, none of them consider the stitch in layout graph.
From Fig. \ref{fig:tplec_speedup_nostitch} we can see that with speed-up techniques (bridge computation and end-cut pre-selection), the runtime can be reduced by around 60\%.

\begin{figure}[htb]
  \centering
  \includegraphics[width=0.84\textwidth]{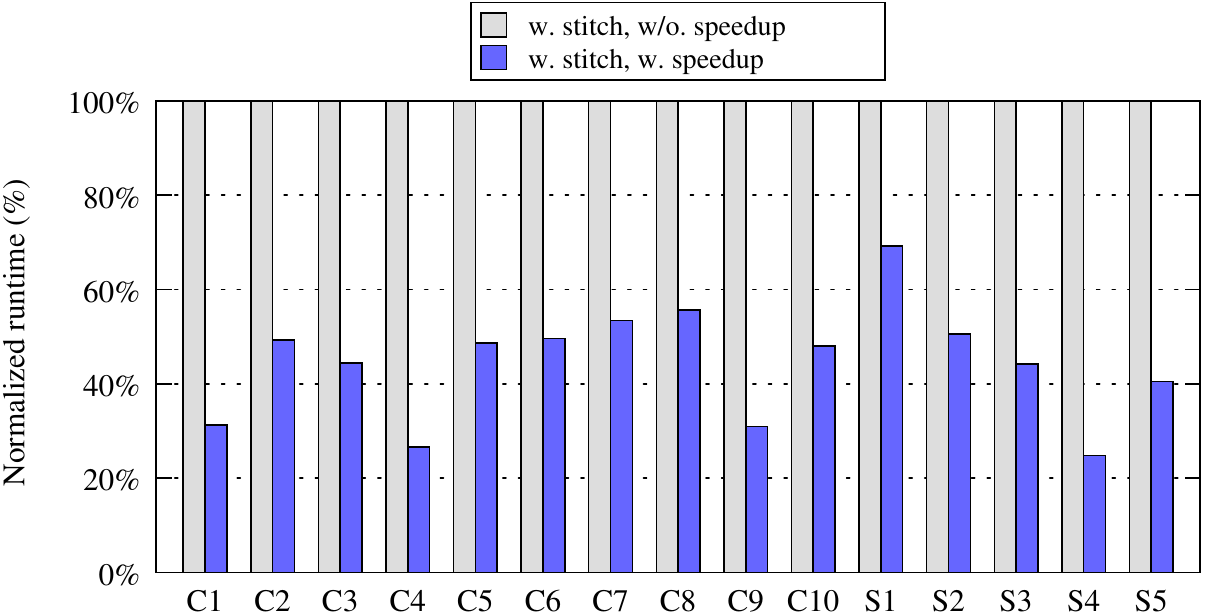}
  \caption{Effectiveness of Speed-up techniques when stitch is introduced.}
  \label{fig:tplec_speedup_stitch}
\end{figure}

Fig. \ref{fig:tplec_speedup_stitch} demonstrates the similar effectiveness of speed-up techniques between ``\textbf{w. stitch w. speedup}'' and ``\textbf{w/o. stitch w. speedup}''.
Here stitch candidates are introduced in layout graph.
We can see that for these two ILP formulation the bridge computation and the end-cut pre-selection can reduce runtime by around 56\%.

\subsection{Conflict Analysis}

\begin{figure}[htb]
  \centering
  \includegraphics[width=0.94\textwidth]{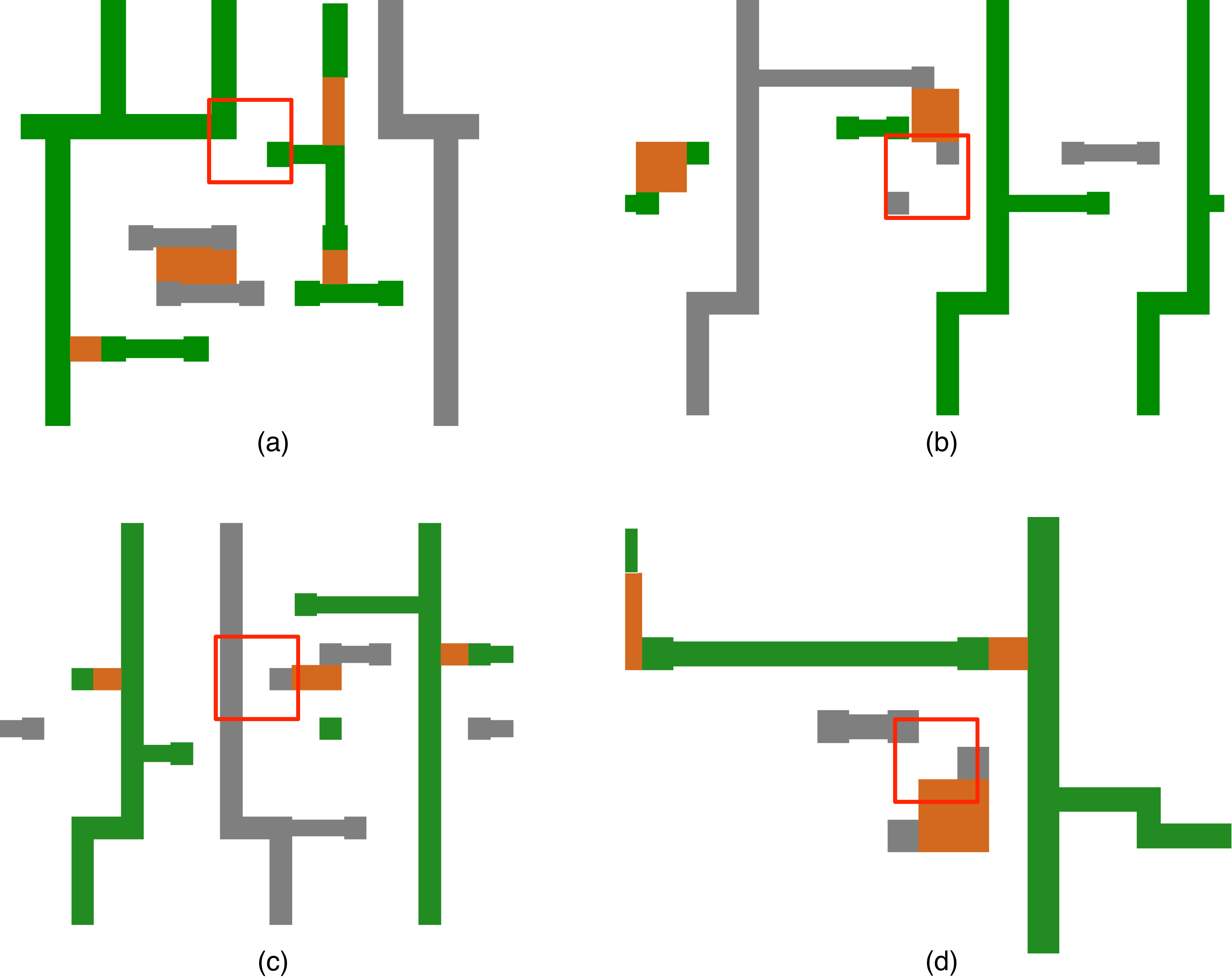}
  \caption{Conflict examples in decomposed results.
  (a)(c) Conflicts because no additional end-cut can be inserted due to the existing neighboring end-cuts.
  (b)(d) Conflicts because no end-cut candidates between irregular vias.
  }
  \label{fig:tplec_conflicts}
\end{figure}

Fig. \ref{fig:tplec_conflicts} shows four conflict examples in decomposed layout, where conflict pairs are labeled with red arrows.
We can observe that some conflicts (see Fig. \ref{fig:tplec_conflicts} (a) and (c)) are introduced due to the end-cuts existing in neighboring.
For these two cases, the possible reason is the patterns are irregular, therefore some end-cuts that close to each other cannot be merged into a larger one.
We can also observe some conflicts (see Fig. \ref{fig:tplec_conflicts} (b) and (d)) come from via shapes.
For these two cases, one possible reason is that it is hard to find end-cut candidates around via, comparing with long wires.

\section{Conclusions}
\label{sec:tplec_conclu}

In this paper we have proposed an improved framework and algorithms to solve the LELE-EC layout decomposition problem.
New end-cut candidates are generated considering potential hotspots.
The layout decomposition is formulated as an integer linear programming (ILP).
The experimental results show the effectiveness of our algorithms.
It shall be noted that our framework is very generic that it can provide high quality solutions to both uni-directional and bi-directional layout patterns.
However, if all the layout patterns are uni-directional, there might be some faster solutions.
Since end-cutting can provide better printability than traditional LELELE process,
we expect to see more works on the LELE-EC layout decomposition and LELE-EC aware design.

\bibliographystyle{spiejour}

\end{spacing}
\end{document}